\documentclass[aps,prl,reprint,superscriptaddress,amsfonts,amsmath,amssymb]{revtex4-1}

\usepackage{graphicx}
\usepackage{bm}
\usepackage{epstopdf}

\begin{document}

\title{Quantum point spread function for imaging trapped few-body systems with a quantum gas microscope}

\author{Sven Kr\"onke}
	\affiliation{Zentrum f\"ur Optische Quantentechnologien, Universit\"at
Hamburg, Luruper Chaussee 149, 22761 Hamburg, Germany}
\author{Maxim Pyzh}
	\email{mpyzh@physnet.uni-hamburg.de}
	\affiliation{Zentrum f\"ur Optische Quantentechnologien, Universit\"at
Hamburg, Luruper Chaussee 149, 22761 Hamburg, Germany}
\author{Christof Weitenberg}
	\affiliation{Institut f\"ur Laserphysik, Universit\"at Hamburg, 
Luruper Chaussee 149, 22761 Hamburg, Germany}
	\affiliation{The Hamburg Centre for Ultrafast Imaging, Universit\"at 
Hamburg, Luruper Chaussee 149, 22761 Hamburg, Germany}
\author{Peter Schmelcher}
	\email{pschmelc@physnet.uni-hamburg.de}
	\affiliation{Zentrum f\"ur Optische Quantentechnologien, Universit\"at 
Hamburg, Luruper Chaussee 149, 22761 Hamburg, Germany}
	\affiliation{The Hamburg Centre for Ultrafast Imaging, Universit\"at 
Hamburg, Luruper Chaussee 149, 22761 Hamburg, Germany}

\begin{abstract}
Quantum gas microscopes, which image the atomic occupations in an optical lattice, 
have opened a new avenue to the exploration of many-body lattice systems. 
Imaging trapped systems after freezing the density distribution by ramping up a pinning lattice leads, 
however, to a distortion of the original density distribution, 
especially when its structures are on the scale of the pinning lattice spacing.
We show that this dynamics can be described by a filter, 
which we call in analogy to classical optics a quantum point spread function. 
Using a machine learning approach, 
we demonstrate via several experimentally relevant setups that a suitable deconvolution allows 
for the reconstruction of the original density distribution. 
These findings are both of fundamental interest for the theory of imaging 
and of immediate importance for current quantum gas experiments. 
\end{abstract}

\maketitle

{\it Introduction.}
Imaging with high resolution is a cornerstone for understanding the structure, dynamics and functionality of matter \cite{Binnig1987,Huang2009,Fernandez-Leiro2016}. In the field of ultracold atoms, quantum gas microscopes have opened new avenues for studying lattice systems \cite{Bakr2010, Sherson2010, Greif2016, Cheuk2016} and led to remarkable progress and insights, such as density correlations and string order \cite{Endres2011}, long-range anti-ferromagnetic correlations \cite{Mazurenko2017} or entanglement growth \cite{Islam2015} in Mott insulators. Naturally, it is of equal interest to study trapped, i.e. \textit{non-lattice systems}, where imaging with single-atom sensitivity is also vital for exploring beyond mean-field physics, i.e. for probing correlation effects \cite{Wenz2013}. Single-atom resolved imaging in free space has been demonstrated for metastable helium atoms, which can be detected using a multi-channel plate with a typical resolution of 60\,$\mu$m \cite{Cayla2017}, and recently for lithium atoms using a short fluorescence pulse, where the position spread due to scattering recoils can be reduced to 4\,$\mu$m \cite{Bergschneider2018}. In order to reach sub-micron resolution, the positions of the atoms have to be frozen by ramping up a pinning lattice before the fluorescence imaging and detection of the atoms takes place. Such a capture of atoms in a pinning lattice was demonstrated starting from a larger scale lattice \cite{Omran2015} or a larger scale continuous system \cite{Hohmann2017}, but freezing and measuring of density structures on the scale of the pinning lattice spacing was so far not considered and achieved. 

Alternative schemes to reach sub-lattice resolution of quantum gases, inspired by related imaging techniques in other fields, have been proposed. Stimulated emission depletion microscopy~\cite{Hell1994}, which breaks the diffraction limit set by the imaging wavelength, can be adapted to quantum gases using the position-dependent dark state of a Lambda-system~\cite{Gorshkov2008}. A scanning tunneling microscope could be realized by coupling to a single ion~\cite{Kollath2007} or by using dispersive couplings to a cavity~\cite{Yang2018}. Momentum mapping in combination with phase retrieval should allow imaging with 1--2 orders of magnitude better than the lattice spacing~\cite{Luhmann2015}. Finally, scanning electron microscopy was successfully applied to quantum gases reaching a resolution below 150\,nm~\cite{Gericke2008}. However, the combination of sub-micron resolution and single-atom sensitivity has so far only been achieved by fluorescence imaging in a pinning lattice.

\begin{figure}[b]
	\includegraphics[width=1.0\linewidth]{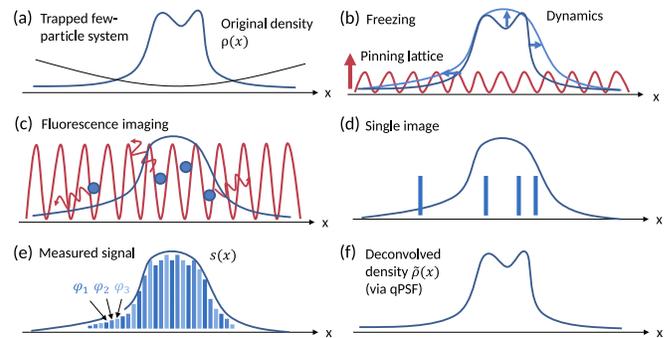}
	\caption{
	(Color online) Protocol for imaging trapped few-body systems with a quantum gas microscope. (a) A few-particle system with one-body density $\rho(x)$ (blue line) is prepared in a trap (black line). (b) The positions of the atoms are frozen by switching off the trap and ramping up a pinning lattice (red line). Due to the dynamics during the ramp, the density is distorted. (c) The positions of the atoms are detected via fluorescence imaging. (d) Individual possible measurement outcomes in a single image. (e) The measurement signal is obtained as the average over many realizations, but it contains the deformation due to the ramp dynamics. By repeatedly preparing a realization of the system and freezing with different phases of the pinning lattice $\varphi$, the density after the ramp can be sampled with a resolution below the lattice spacing $a_{\rm l}$. (f) By deconvolution with the quantum point spread function (qPSF), the original density can be recovered. All sub-figures show sketches.}
	\label{fig:scheme}
\end{figure}

Here, we propose to perform repeated measurements with shifted positions of the pinning lattice relative to the trapped physical system, such that a resolution below the lattice spacing becomes possible and we provide an in-depth analysis of this protocol.
We show that the density structures on the scale of the lattice spacing will be distorted due to the dynamics taking place during the ramp-up of the pinning lattice. The lattice ramp has to be sufficiently fast to avoid an adiabatic loading of the ground state of the lattice, but sufficiently slow to avoid projections onto very high bands, where the atom positions are not frozen due to large tunneling rates. The proposed scenario is illustrated in Fig.\,\ref{fig:scheme}. We show that the distortions during the ramp-up can be captured by a quantum point spread function (qPSF). Using deconvolution techniques, these distortions can be removed, which enhances the resolution of the overall measurement sequence. We find that the deconvolution is both relevant and effective for density structures on the scale of the lattice spacing and provides a sub-wavelength resolution. Our approach and technique suggests itself for immediate applications, because a tight confinement and resulting small structures of the original trapped system allow for strongly interacting quantum systems, while the spacing of the pinning lattice is fixed to typically 0.5\,$\mu$m by the optical wavelength of the interfering laser beams.

{\it Quantum point spread function.} 
We first derive the qPSF for the measurement of a single particle in the pre-measurement state $|\phi\rangle$
\footnote{Our results can be straightforwardly adapted to mixed pre-measurement states.} and then extend the concept to many-body systems.
The measurement is modeled as a two-step process: the ramp-up of the pinning lattice and the read-out of the state occupations. In the following, we keep the phase off-set of the pinning lattice $\varphi$ fixed and thereafter vary it for resolving fine density structures.
During the ramp-up, we assume that all external potentials but the pinning lattice are either switched off or negligible such that the quantum dynamics is governed by the Hamiltonian
\begin{equation}
	\hat h_\varphi(t) =\frac{\hat p^2}{2m}+V(t)\,\sin^2\left(k_{\rm l} \hat{x}-\varphi\right).
	\label{eq:lattice_hamilt_with_units}
\end{equation}
Here, the lattice depth $V(t)$ is ramped up from zero to its maximal value $V_{\rm f}$ within the time-scale $T_{\rm f}$ by using a tanh-like ramping protocol and $k_{\rm l}$ denotes the pinning lattice wavenumber corresponding to the lattice spacing $a_{\rm l}$. The lattice sets the recoil energy $E_{\rm r}=\hbar^2 k_{\rm l}^2 /(2m)$ as typical energy scale. Directly after ramping up the lattice, the system is in the state $\hat U_\varphi|\phi\rangle$ with  $\hat{U}_\varphi=\hat{T}\exp(-i/\hbar\int_0^{T_{\rm f}}{\rm d}\tau\,\hat h_\varphi(\tau))$ 
and $\hat{T}$ denoting the chronological time-ordering operator.

The occupation of the site $i$ is then read out via fluorescence imaging, which we describe within the established framework of measurement operators $\hat R_{i;\varphi}$ and positive operator-valued measures $\hat R^\dagger_{i;\varphi}\hat R_{i;\varphi}$ \cite{Wiseman2014}. Being only interested in the probability for finding the particle at site $i$ given the phase off-set $\varphi$
\begin{equation}
	p_{i;\varphi}=\langle\phi|\hat U_\varphi^\dagger \hat R^\dagger_{i;\varphi}\hat R_{i;\varphi} \hat U_\varphi|\phi\rangle,
\end{equation}
we have to specify the operator $\hat R^\dagger_{i;\varphi}\hat R_{i;\varphi}$. For this purpose, we assume that a particle that ends up in the Wannier state $|w^\alpha_{i;\varphi}\rangle$ of the pinning lattice Hamiltonian $\hat h_\varphi(T_{\rm f})$ after the ramp-up, where $\alpha$ denotes the band index, is measured with the detection efficiency $\eta_\alpha\in [0,1]$, which can be modeled by
\begin{equation}
	\hat R^\dagger_{i;\varphi} \hat R_{i;\varphi}= \sum_\alpha\,\eta_\alpha\,|w^\alpha_{i;\varphi}\rangle\!\langle w^\alpha_{i;\varphi}|.
	\label{eq:measurement}
\end{equation}
Here, a high detection efficiency $\eta_\alpha$ is ensured, if the tunneling rate of the band $J_{\alpha}$ is small compared to the imaging time scales. As $J_{\alpha}$ increases very rapidly for higher bands $\alpha$, we can approximate the efficiencies by a step function, i.e. $\eta_{\alpha}=1$ for a finite number of 'non-tunneling bands' and $\eta_{\alpha}=0$ for all higher bands. Then the operator $\hat R_{i;\varphi}^\dagger\hat R_{i;\varphi}$ becomes a projector. Atoms in higher bands or continuum states 
\footnote{We note that $\sum_i \hat{R}^\dagger_{i;\varphi} \hat{R}_{i;\varphi}\neq \openone$ due to continuum states and detection efficiencies $\eta_{\alpha}<1$. Yet our measurement model can be easily extended to a proper positive operator-valued measure by associating the operator $\openone-\sum_i\hat R^\dagger_{i;\varphi} \hat R_{i;\varphi}$ with the loss measurement outcomes.} lead to loss and the lattice ramp has to be chosen such that this loss remains small. Deep lattices and not-too-fast ramps keep this loss negligible. Finally, the measurement signal $s(x)$ is obtained by averaging over the pinning lattice shifts $\varphi$.

\begin{figure}[t]
	\includegraphics[width=1.0\linewidth]{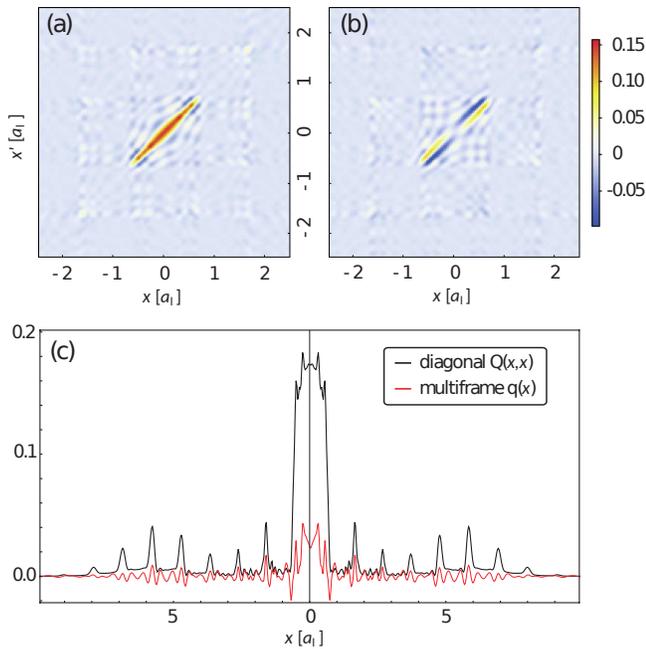}
	\caption{(Color online) Filter for image deconvolution. (a) Real part and 
	(b) imaginary part of the quantum point spread function $Q(x,x')$. 
	(c) Comparison of the filter from diagonal approximation (black) and multi-frame filter (red) 
	learned from physical examples (see text). 
	Ramping parameters are $V_{\rm f}=200 E_{\rm r}$ and $T_{\rm f}=\hbar/E_{\rm r}$.}
	\label{fig:filter}
\end{figure}

In order to define the qPSF we consider the analogy to classical optics, where the exact ``object'' density $\rho(x)$ becomes blurred in the image plane via the point spread function $f(x)$ according to the convolution $s(x)=(f\ast\rho)(x) = \int {\rm d}y\,\rho(y)f(x-y)$, where $s(x)$ denotes the signal in the imaging plane. Given this relationship and the precise form of $f$, there are various deblurring techniques for (approximately) restoring $\rho(x)$. Our aim here is to reformulate the probability  $p_{i;\varphi}$
as a convolution to define qPSF for our imaging protocol. 
By means of the translation symmetry $\hat U_\varphi^\dagger \hat R^\dagger_{i;\varphi}\hat R_{i;\varphi} \hat U_\varphi
=\hat T_{ia_l+\varphi}\hat U_0^\dagger \hat R^\dagger_{0;0}\hat R_{0;0} \hat U_0\hat T_{ia_l+\varphi}^\dagger$
with the translation operator $\hat T_{z}=\exp(-i z\hat p/\hbar)$, we arrive at our central result:
\begin{equation}
	 p_{i;\varphi}=\int\int{\rm d}x{\rm d}y\,\phi^*(x)\;Q(z-x,z-y)\;\phi(y)\Big|_{z=i a_{\rm l}+\varphi},
	 \label{QPSF1}
\end{equation}
where the kernel $Q(x,y)=\langle-x|\hat U_0^\dagger \hat R^\dagger_{0;0}\hat R_{0;0} \hat U_0|-y\rangle$ describes both the quantum dynamics during the ramp-up of the pinning lattice
and the subsequent fluorescence imaging. Thus, we find that the probability
of detecting the particle at site $i$ given the pinning lattice offset $\varphi$
is provided by the diagonal of the two-dimensional ($2D$) convolution of $\phi^*(x)\phi(y)$
with $Q(x,y)$, which we therefore name quantum point spread function (see Fig.\,\ref{fig:filter}).
Eq.\,\eqref{QPSF1} moreover shows that the probability $p_{i;\varphi}$ can be expressed by
a continuous function $s(z)$ evaluated at discrete positions, $p_{i;\varphi}=s(i a_{\rm l}+\varphi)$.
By repeating the experiment for various offsets $\varphi$ one effectively samples
this pseudo-probability $s(z)$, which shall be called \textit{signal} in the following.
For practical calculations, one can spectrally decompose the qPSF and finds that the signal $s(z)$ equals an incoherent superposition of $1D$ convolutions of the pre-measurement state with the back-propagated Wannier states
$|\chi_\alpha\rangle=\hat\pi\hat U^\dagger_{0}|w^\alpha_{0;0}\rangle$:
\begin{equation}
	s(z)=\sum_\alpha\eta_\alpha\,\Big|\int{\rm d}x \chi^*_\alpha(z-x)\phi(x)\Big|^2,
\end{equation}
where $\hat\pi$ denotes the parity operator. As a side remark, the quantum dynamics during the ramp is non-adiabatic such that $\hat\pi|\chi_\alpha\rangle$ does not coincide with the corresponding Wannier state of the shallower lattices of the ramp.

In order to extend the qPSF to many-body systems, which is accomplished in the supplementary material \cite{SupMat}, we make the following two assumptions: first, we assume that all interactions are switched off, e.g., via a Feshbach resonance before ramping up the pinning lattice and reading off the site occupations. Second, the atomic density should be small enough such that light-assisted collisions can be neglected during the imaging \cite{Bakr2010,Sherson2010}. Then, we can factorize the qPSF describing the statistics of single-shot measurements upon an $N$-body ensemble by $N$ single-particle qPSF \cite {SupMat}. Moreover, the ensemble average over many such single-shots results in the signal $s(z)=\int\int{\rm d}x{\rm d}y\,\rho_1(x,y)\;Q(z-x,z-y)$ with $\rho_1(x,y)$ denoting the pre-measurement reduced one-body density matrix, if the population of unobserved pinning-lattice bands and continuum states after the ramp-up is negligible \cite{SupMat}. We note that the spatial dimensionality does not play a role and the framework equally applies to higher spatial dimensions.

{\it Deconvolution.}
Inverting the relationship Eq.\,\eqref{QPSF1} is quite a difficult task: 
deconvolution in general is an ill-posed problem and, moreover, we have to cope with the intriguing situation that the measurement signal $s(z)$ constitutes only the diagonal of the $2D$ convolution $(Q*\rho_1)(z,z')$
\footnote{For (quasi) one-dimensional systems.}. 
By scanning over different lattice ramps we find
for suitably chosen $V_{\rm f}$ and $T_{\rm f}$ that the real part of the qPSF $Q(x,x')$ acquires a dominant diagonal pattern with a fast decay of the off-diagonal elements, while the imaginary part is significantly smaller [see Fig.\,\ref{fig:filter}(a),(b)]. This motivates us to express the signal $s(z)$ as a $1D$ convolution of the one-body density $\rho(x)=\rho_1(x,x)$ with some yet unknown $1D$ filter $q(x)$:
\begin{equation}
	s(z)\approx 
	\int{\rm d}x\,q(z-x)\,\rho(x)=(q*\rho)(z).
	\label{eq:1D_approx}
\end{equation}
However, it is a priori not clear, whether such a filter, which is independent of the underlying density, exists and if it does,
how to obtain it. Yet, if one has found such a filter, Eq.\,\eqref{eq:1D_approx} allows for applying established deconvolution algorithms for obtaining the pre-measurement density $\rho(x)$. Making the most obvious choice by taking the diagonal of the qPSF, $q(x)=Q(x,x)$, turns out to be numerically unstable and inaccurate. To compensate for the complexity of the $2D$ convolution, this diagonal needs to be readjusted. To this end we call upon a machine learning approach.

Inspired by the multi-frame deconvolution technique, which is applied, e.g., in astronomy \cite{onlineMFBD}, we pursue the following machine learning approach to learn the unknown filter $q(x)$. Our training set consists of a small number $n_{\rm t}$ of one- and many-body states with known (reduced) one-body density matrix $\rho_1^{(k)}(x,y)$, $k=1,...,n_{\rm t}$ \cite{SupMat}. For each training sample, we calculate the corresponding measurement signal $s^{(k)}(z)$ by the full $2D$ convolution with the exact qPSF $Q(x,y)$ 
\footnote{Recall that $Q(x,y)$ depends only on the ramp parameters and not on the physical sample.}. Then, we determine the best $1D$ filter $q(z)$ by minimizing the mean-squared error on the training set $\sum_{k=1}^{n_{\rm t}}\int{\rm d}z\,[s^{(k)}(z)-(q*\rho^{(k)})(z)]^2$ via batch gradient-descent with line-search \cite{SupMat} [see Fig.\,\ref{fig:filter}(c)]. 
We finally apply a classical deconvolution algorithm \cite{SupMat} to Eq.\,\eqref{eq:1D_approx} 
for several unseen cases $s(z)$ to reconstruct the underlying pre-measurement density $\rho(x)$.

\begin{figure*}
	\includegraphics[width=1.0\linewidth]{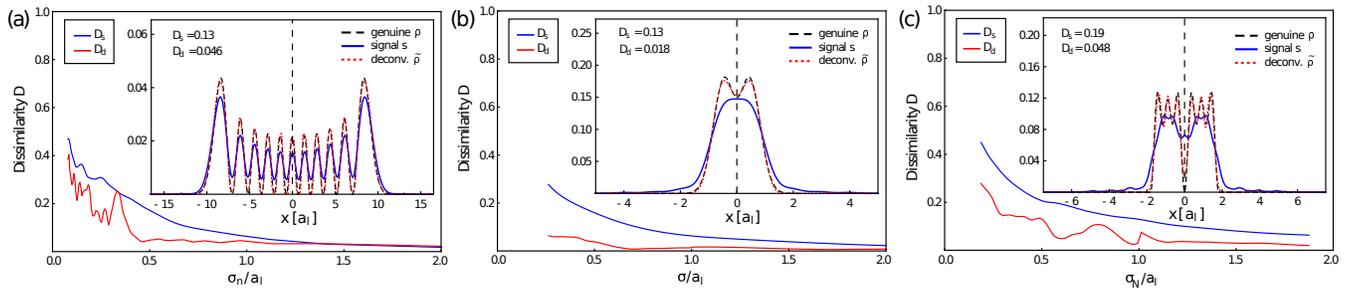}
	\caption{(Color online) Three different physical example situations demonstrating the performance of the multi-frame filter. (a) $n=10$ excited eigenstate of the harmonic oscillator for a varying trapping frequency (inset shows $\omega=2\pi\times100\,$Hz). (b) Ground state of two indistinguishable bosons with infinite repulsion in a harmonic trap with varying trapping frequencies (inset shows $\omega=2\pi\times1.14\,$kHz). (c) Fermi polaron, i.e. few fermions in a box with a $\delta$-potential at the origin for $N=6$ particles and varying box lengths $L_{\rm b}$. All examples are for the case of $^{87}$Rb atoms in a lattice with $a_{\rm l}=532\,$nm. The insets show the genuine density (black dashed line) along with the measurement signal after freezing the distribution with a lattice ramp to $V_{\rm f}=200 E_{\rm r}$ in $T_{\rm f}=\hbar/E_{\rm r}$ (blue solid line) and the deconvolved signal using the multi-frame filter (red dotted line). For the chosen examples, the measurement signal has clear distortions from the dynamics during the ramp-up, which are, however, removed by the deconvolution. The main panels show the deviations [Eq.\,\eqref{eq:deviations}] of the measurement signal $D_{\rm s}$ (blue solid line) and the reconstructed density $D_{\rm d}$ (red solid line) from the genuine singe-particle density as a function of the structure size (see \cite{SupMat} for definitions).}
	\label{fig:examples}
\end{figure*}

{\it Applications.} 
We showcase the performance of our qPSF approach and deconvolution strategy using three physical example setups (see Fig.\,\ref{fig:examples}): excited harmonic oscillator eigenstates featuring a rapidly oscillating density, two identical bosons with infinite repulsion in a harmonic trap~\cite{Busch1998} and a Fermi polaron. For details on the implementation of these systems see \cite{SupMat}. The examples are chosen to cover a broad range of different situations: single particle, weakly- and highly-correlated few-body physics. 

The deconvolution uses a multi-frame filter $q$ [see Fig.\,2(c)] 
trained with a random selection of sample densities of the harmonic oscillator example 
and a dark soliton in a small BEC as additional training example (see \cite{SupMat}).
We find that applying it to the unknown signals from the two-Boson and Fermi polaron problem, yields very good results, emphasizing the power of the method. We stress that we learn the multi-frame filter from single-particle and mean-field cases and then apply it to the unseen situations, which involve correlated many-body states.

The insets in Fig.\,\ref{fig:examples} show the genuine single-particle density $\rho(x)$, the measurement signal $s(x)$ and the deconvolved signal $\tilde \rho(x)$ for different physical examples. In all cases, the structure of the genuine density is washed out in the measurement signals, but almost completely recovered by the deconvolution. In particular, we recover all of the many oscillations for the harmonic oscillator with their full original contrast [Fig.\,\ref{fig:examples}(a)]. In the two-Boson example, it is striking that the deconvolution successfully reproduces the original density although the two humps have merged into a single one in the measurement signal [Fig.\,\ref{fig:examples}(b)]. 
In the Fermi polaron example, both the sharp dip in the center of the trap 
and the Friedel oscillations around it are fully recovered in the deconvolved signal, 
although they seemed to be lost in the measurement signal [Fig.\,\ref{fig:examples}(c)]. 
These examples showcase the power of the deconvolution method using the qPSF.

To judge on the quality, we introduce a dissimilarity measure between two normalized functions $g$ and $h$ as
\begin{equation}
	D(g,h) = \frac{1}{2}  || g - h||_1 = \frac{1}{2}  \int dx \ |g(x)-h(x)|.
	\label{eq:deviations}
\end{equation}
It takes a value of zero for coinciding functions and increases up to one as the absolute deviation becomes more pronounced. 
Further, we define the dissimilarity between the measurement signal $s$ 
\footnote{$s(z)$ is in general not exactly normalized 
due to the small particle losses induced by the chosen modeling of quantum efficiencies. 
We normalize it to unity for the dissimilarity analysis and the deconvolution procedure.} 
and the genuine density $\rho$ as $D_{\rm s}=D(s,\rho)$
and the dissimilarity between the reconstructed density $\tilde{\rho}$ and the genuine one as $D_{\rm d}=D(\tilde{\rho},\rho)$.
In Fig.\,\ref{fig:examples}, we show how this dissimilarity depends on the typical structure size $\sigma$ of the genuine density. For structures that are large compared to the lattice spacing ($\sigma > 2 a_{\rm l}$), the dissimilarity $D$ is negligible both for the measurement and the deconvolved signal. When the structures are on the scale of the lattice spacing, the measurement signal starts to deviate due to the dynamics during the ramp-up of the pinning lattice. The dissimilarity of the deconvolved signal, however, remains negligible due to the successful deconvolution. Only for structures smaller than about half the lattice spacing ($\sigma < 0.5 a_{\rm l})$, we observe an increase of $D_{\rm d}$, indicating the limitations of the method. Using the deconvolution via the qPSF, we can therefore shift the accessible structure sizes from about $2 a_{\rm l}$ to about $0.5 a_{\rm l}$, which is a significant improvement that is crucial for many physical examples in quantum gas physics.

{\it Outlook.} 
Our work opens the research direction of high-resolution imaging with single-atom sensitivity also for trapped, i.e. non-lattice systems. We propose to apply a pinning lattice for imaging and to sample the reduced one-body density with a resolution below the lattice spacing by performing repeated measurements with shifted positions of the pinning lattice relative to the physically trapped system. We have shown that density distortions resulting from the dynamics during the ramping up of the lattice can be compensated by deconvolution with a quantum point spread function for a wide range of parameters. Our findings are of immediate relevance for ongoing quantum gas microscope experiments. A reliable measurement of small density structures will allow accessing new regimes and imaging of the corresponding physical processes such as the shape of a vortex core taking into account beyond mean-field effects \cite{Barberan2006} or discrete few-body structures in arbitrary traps.
For simplicity, we have focused here on one-dimensional systems, but our framework equally applies to higher spatial dimensions. Further extensions of our work would be the fate of correlation measurements \cite{Schauss2012} and blurring effects in the measurement of the dynamics. Another important aspect is the imaging after release from a driven system, e.g., for producing artificial gauge fields \cite{Dalibard2011}, where switching off the drive can induce further effects. Releasing from lowest Landau levels yields a self-similar expansion of the wave function \cite{Read2003}, which could be used before freezing the distribution.

\begin{acknowledgments}
P. S. and C. W. gratefully acknowledge funding by the Deutsche Forschungsgemeinschaft in the framework of the SFB 925 `Light induced dynamics and control of correlated quantum systems'. 
\end{acknowledgments}

S. K. and M. P. contributed equally to this work.

\bibliographystyle{apsrev4-1}

\setcounter{equation}{0}
\setcounter{figure}{0}
\setcounter{table}{0}
\renewcommand{\theequation}{S\arabic{equation}}
\renewcommand{\thefigure}{S\arabic{figure}}
\renewcommand{\thetable}{S\arabic{table}}

\appendix
\newpage
\onecolumngrid

\section{Supplementary material: Quantum point spread function for imaging trapped few-body systems with a quantum gas microscope}

\subsection{Section A: Extension of the qPSF theory to many-body systems}
In the following, we extend the qPSF theory to many-body systems 
by first inspecting the case of distinguishable particles, then discussing single-shot measurements of indistinguishable particles, and finally deriving the relationship between the ensemble average of such single-shot measurements and few-body correlations.

\subsection{Single-shot measurements of distinguishable particles}
In order to extend the qPSF theory to many-body systems, we make the following assumptions:
(i) As in the single-particle case, we assume that all external traps but the 
pinning lattice are either switched off or negligible during the full measurement protocol.
(ii) Moreover, we assume that the inter-atomic interactions are either switched off
by means of a Feshbach resonance or negligible during the full measurement protocol.
(iii) Finally, we regard the fluorescence imaging of the pinning lattice sites 
to be a pure one-body process, i.e.\
neglect few-body effects such as loss via light-induced collisions  \cite{Bakr2010,Sherson2010}. The latter
approximation is valid, if the pre-measurement atomic density is so low that the
likelihood of finding more than one atom in a pinning-lattice site after ramp-up
is strongly suppressed. 

Under the above assumptions, an $N$-body system in the pure
\footnote{The extension to mixed pre-measurement states is straightforward.} pre-measurement state $|\Psi\rangle$ evolves into
the state $\hat U_\varphi^{(1)}\otimes...\otimes\hat U_\varphi^{(N)}\,|\Psi\rangle$
during the ramp-up of the pinning lattice with given phase off-set $\varphi$. Here
$\hat U_\varphi^{(\kappa)}=\hat{T}\exp[-i/\hbar\int_0^{T_{\rm f}}{\rm d}\tau\,\hat h_\varphi^{(\kappa)}(\tau)]$
denotes the single-particle time-evolution operator acting on the $\kappa$-th particle.

Since the fluorescence imaging is modeled as a single-particle process, we can
directly transfer the positive operator-valued measure for the single-particle
case [see Eq.\,(3) of the main text] to the many-body realm and obtain for the probability to detect the 1st, 2nd, ..., $N$-th particle
in the pinning-lattice site $i_1$, $i_2,$ ..., $i_N$, respectively:
\begin{equation}
	P_\varphi(i_1,...,i_N)=\langle\Psi|
	\hat{M}^{(1)}_{i_1;\varphi}\otimes...\otimes
	\hat{M}^{(N)}_{i_N;\varphi}|\Psi\rangle,
	\label{eq:N-dist_prob1}
\end{equation}
where $\hat M^{(\kappa)}_{i_\kappa;\varphi}\equiv[\hat U_\varphi^\dagger \hat R^\dagger_{i_\kappa;\varphi}\hat R_{i_\kappa;\varphi} \hat U_\varphi]^{(\kappa)}$ (the bracket $[...]^{(\kappa)}$ shall indicate that the whole operator acts on the $\kappa$-th particle).
Note that for a fixed phase the
probability $\sum_{\bf i} P_\varphi(\bf i) \leq 1$ due to the possibility of
detection efficiencies being smaller than one.
Making use of the translation symmetry of $\hat M^{(\kappa)}_{i_\kappa;\varphi}$ as in the single-particle case discussed in detail in the main text, we may express Eq.\,\eqref{eq:N-dist_prob1} as
\begin{equation}
	P_\varphi(i_1,...,i_N)=\int {\rm d}^N\!x\,{\rm d}^N\!y\;\Psi^*({\bf x})\,
	Q_N({\bf z}-{\bf x},{\bf z}-{\bf y})\,\Psi({\bf y})\Big|_{{\bf z}={\bf i} a_{\text{l}}+\varphi},
	\label{eq:N-dist_prob2}
\end{equation}
where the spatial positions of the
$N$ particles are abbreviated as ${\bf x}\equiv(x_1,...,x_N)$
and the integrals are taken w.r.t.\ all particle coordinates, i.e., ${\rm d}^N\!x\equiv
{\rm d}x_1\,{\rm d}x_2...{\rm d}x_N$. Moreover, the pinning-lattice sites, in which the particles are detected, are abbreviated as ${\bf i}\equiv(i_1,...,i_N)$
and $\Psi({\bf x})\equiv\langle x_1,...,x_N|\Psi\rangle$ refers to the 
position representation of the $N$-body pre-measurement state $|\Psi\rangle$.
Finally, the $N$-body qPSF turns out to be the $N$-fold product of the one-body
qPSF, which has been derived for the single-particle case in the main text:
\begin{equation}
	Q_N({\bf x},{\bf y})=\prod_{\kappa=1}^NQ(x_\kappa,y_\kappa),
\end{equation}
where $Q(x_\kappa,y_\kappa)=\langle-x_\kappa|\hat M^{(\kappa)}_{0;0}|-y_\kappa\rangle$.
Thereby, we obtain the $N$-particle post-measurement distribution $P_\varphi(i_1,...,i_N)$ for a given pinning lattice
phase off-set $\varphi$ by evaluating the signal function
\begin{equation}
	S({\bf z}) = \int {\rm d}^N\!x\,{\rm d}^N\!y\;\Psi^*({\bf x})\,
	Q_N({\bf z}-{\bf x},{\bf z}-{\bf y})\,\Psi({\bf y})
\end{equation}
at the discrete positions ${\bf z}={\bf i} a_{\text{l}}+\varphi$, i.e.,
$P_\varphi({\bf i}) = S({\bf i} a_{\text{l}}+\varphi)$. Repeating the $N$-body
measurement for various phase off-sets $\varphi$ effectively means sampling
from the pseudo-probability $S({\bf z})$.

\subsection{Single-shot measurements of indistinguishable particles}
Next, we concentrate on the special case of $N$ indistinguishable particles. In this case, the outcome of a single-shot measurement is a pinning lattice occupation-number histogram $(n_1,...,n_L)\equiv{\bf n}$, where $n_i$ denotes the number of particles found in the $i$-th lattice site, $i=1,...,L$ and $L$ refers to the number of lattice sites. Here, we have in particular few-body situations in mind, where one can easily probe the full distribution of the $N$ particles in the pinning lattice. 

Obviously, the $N$-body qPSF $Q_N({\bf x},{\bf y})$ remains invariant under simultaneous permutation of the particle labels in both
${\bf x}$ and ${\bf y}$. Given a system of indistinguishable particles, 
the probability Eq.\,\eqref{eq:N-dist_prob2}
does not depend on the concrete particle labeling but only on how many
particles are found in a certain site. Taking this combinatorically into account,
one finds for the probability of the histogram ${\bf n}$ for a given 
phase off-set $\varphi$
\begin{equation}
	\bar P_\varphi(n_1,...,n_L)=\frac{N!}{\prod_{i=1}^Ln_i!}\,P_\varphi({\bf i_{\bf n}}),
	\label{eq:N-indist_prob}
\end{equation}
where ${\bf i}_{\bf n}$ denotes some $N$-dimensional lattice-site index 
vector, which features $n_r$-times the entry $r$ with $r=1,...,L$. We remark
that while Eq.\,\eqref{eq:N-indist_prob} describes the (within the considered 
measurement model) correct probability of detecting the histogram ${\bf n}$ given the number of particles $N$ and the phase off-set $\varphi$, these probabilities do not sum up to unity in general when considering all conceivable histograms ${\bf n}$ with $N$ particles.
In fact, the probability for not detecting all $N$ particles in the pinning lattice
due to the occupation of higher bands with detection efficiencies smaller than unity
or continuum states after ramp-up [see Eq.\,(3) of the main text] reads
$1-\sum_{{\bf n}|N}\bar P_\varphi({\bf n})$, where the sum $\sum_{{\bf n}|N}$ runs over all histograms
${\bf n}$ with $\sum_{i=1}^L n_i=N$.

\subsection{Ensemble averages over single-shot measurements and few-particle correlations}
Having taken many single-shot measurements of identical copies of the many-body system,
one may evaluate the corresponding ensemble average of certain $n$-particle observables.
In classical absorption imaging of atomic samples for instance,
one obtains the reduced one-body density by averaging the spatial particle number distributions
of many single-shot measurements. Density-density correlations can be inferred from absorption images by averaging 
the product of occupation-number fluctuations at two spatial positions over many single-shot measurements. Here,
we stress that in classical absorption imaging the average of an $n$-particle quantity over many single-shot measurements
is directly connected to the pre-measurement reduced $n$-body density matrix, whereas 
in the case of the quantum gas microscope measurement protocol pursued in this work, 
this relationship is more complicated in general and shall be derived here.

First, let us derive the probability $p^{(1)}_\varphi(r)$ 
to find an atom in the pinning-lattice site
$r$ when averaging over many single-shot measurements with the same phase off-set 
$\varphi$, i.e., many different
histograms ${\bf n}$ distributed according to $\bar P_\varphi({\bf n})$. Using
Eq.\,\eqref{eq:N-indist_prob}, we find
\begin{align}
	p^{(1)}_\varphi(r) &= \sum_{{\bf n}|N}\frac{n_r}{N}\bar P_\varphi({\bf n})\\\nonumber
	&=\sum_{{\bf n}|N}\frac{(N-1)!}{\prod_{i\neq r}n_i!}\,\frac{n_r}{n_r!}\,
	P_\varphi({\bf i}_{\bf n}).
\end{align}
Apparently, only histograms ${\bf n}$ with $n_r>0$ contribute to $p^{(1)}_\varphi(r)$. Substituting
${\bf n}={\bf m}+{\bf e_r}$, where ${\bf m}$ denotes an arbitrary $(N-1)$-particle
histogram and ${\bf e_r}$ an occupation number vector with all components zero except for the
$r$-th one being set to unity, one obtains
\begin{equation}
	p^{(1)}_\varphi(r) = \sum_{{\bf m}|N-1}\frac{(N-1)!}{\prod_{i}m_i!}
	P_\varphi({\bf i}_{\bf m+e_r}).
\end{equation}
Next, we rewrite the summation over $(N-1)$-particle
histograms as a summation over $N-1$ lattice site indices
\begin{equation}
	p^{(1)}_\varphi(r) = 
	\sum_{i_2,...,i_N=1}^L
	P_\varphi(r,i_2,...,i_N).
\end{equation}
Abbreviating, $\hat K_\varphi^{(\kappa)}\equiv\sum_{i=1}^L\hat M_{i;\varphi}^{(\kappa)}$,
we finally obtain
\begin{equation}
	p^{(1)}_\varphi(r)=\langle\Psi|\hat M^{(1)}_{r;\varphi}\otimes\hat
	K_\varphi^{(2)}\otimes...\otimes\hat K_\varphi^{(N)}
	|\Psi\rangle.
\end{equation}
Since particles in higher bands or continuum states after the pinning lattice 
ramp-up are not detected, $\hat K_\varphi^{(\kappa)}\neq\openone$ and thus 
$p^{(1)}_\varphi(r)$ is {\it not} given as the expectation value of a one-body
observable. So the one-particle quantity $p^{(1)}_\varphi(r)$ may depend 
on up to $N$-particle corrections and cannot be represented as the trace
of a one-particle observable times the pre-measurement reduced one-body density
operator in general, which is in contrast to the case of absorption imaging.

Our simulations in the main text, however, show that the probability 
to populate higher bands or continuum states by the pinning lattice ramp-up 
is negligibly small for suitably chosen 
experimental settings 
(see the discussion on the impact of higher bands on the qPSF in Section B as well as Table \ref{tab:losses}).
Under these circumstances, $\hat K_\varphi^{(\kappa)}$ effectively acts as
the identity operator on the $\kappa$-th particle in $|\Psi\rangle$ and we obtain
the relation 
\begin{equation}
	p^{(1)}_\varphi(r) = {\rm tr}\big(\hat M^{(1)}_{r;\varphi}\,\hat\rho_1\big)
	=\int{\rm d}x{\rm d}y\,Q(z-x,z-y)\,\rho_1(x,y)\Big|_{z=r a_{\text{l}}+\varphi}, 
\end{equation}
where $\rho_1(x,y)=\langle x|\hat \rho_1|y\rangle$ denotes the 
position representation of the pre-measurement reduced one-body density operator $\hat\rho_1$,
which one obtains from the pre-measurement many-body state by a partial trace over all but one particle, $\hat\rho_1={\rm tr}_1\big(|\Psi\rangle\!\langle\Psi|\big)$.

Similarly, one can derive the corresponding expressions for the ensemble average of
an $n$-particle quantity with $n>1$ over many single-shot measurements. Here, we only explicate this relationship for the case $n=2$, i.e., the
probability $p^{(2)}_\varphi(r_1,r_2)$ to detect a particle at site $r_1$ and another particle in site
$r_2$ in the ensemble average:
\begin{equation}
	p^{(2)}_\varphi(r_1,r_2)= \sum_{{\bf n}|N}\frac{n_{r_1}}{N}\frac{n_{r_2}}{N-1}
	\bar P_\varphi({\bf n}).
\end{equation}
Similar manipulations as above can be applied and one arrives at
\begin{align}
	p^{(2)}_\varphi(r_1,r_2)&=\sum_{i_3,...,i_N=1}^L
	P_\varphi(r_1,r_2,i_3,...,i_N)\\\nonumber
	&=\langle\Psi|\hat M^{(1)}_{r_1;\varphi}\otimes\hat M^{(2)}_{r_2;\varphi}\otimes\hat
	K_\varphi^{(3)}\otimes...\otimes\hat K_\varphi^{(N)}
	|\Psi\rangle.
\end{align}
If the population of higher bands and continuum states after the
pinning-lattice ramp-up may be neglected, we end up with
\begin{align}
	p^{(2)}_\varphi(r_1,r_2) &= {\rm tr}\big(\hat M^{(1)}_{r_1;\varphi}\otimes
	\hat{M}^{(2)}_{r_2;\varphi}\,\hat\rho_2\big)\\\nonumber
	&=\int{\rm d}^2x\,{\rm d}^2y\,Q_2({\bf z}-{\bf x},{\bf z}-{\bf y})\,\rho_2({\bf x},{\bf y})\Big|_{{\bf z}=
	{\bf r} a_{\text{l}}+\varphi}, 
\end{align}
where ${\bf r}=(r_1,r_2)$ and $\rho_2({\bf x},{\bf y})$
denotes the position representation of the  pre-measurement reduced two-body density operator $\hat\rho_2$,
which one obtains by a partial trace over all but two particles, $\hat\rho_2=
{\rm tr}_2\big(|\Psi\rangle\!\langle\Psi|\big)$.

\subsection{Section B: Numerical procedure to obtain the quantum point spread function}
According to Eq.\,(4) of the main text the qPSF is an operator 
$\hat{Q}=\sum_\alpha\,\eta_\alpha\,|\chi_\alpha\rangle\!\langle \chi_\alpha|$  with
$|\chi_\alpha \rangle=\hat{\pi} \hat U_0^\dagger |w^\alpha_{0;0}\rangle$.
Therefore, we have to calculate the Wannier state $|w^\alpha_{0;0}\rangle$ of the
band $\alpha$ at site $i=0$ for the pinning lattice with the final potential depth $V_{\rm f}$
and phase off-set $\varphi=0$ and propagate it with the time-evolution operator
$\hat U_0^\dagger=\hat{T}\exp(-i/\hbar\int_{T_{\rm f}}^0{\rm d}\tau\,\hat h_0(\tau))$,
which describes the lowering of the pinning lattice from $V_f$ to zero depth. 
Finally, the parity operator $\hat{\pi}$ is applied 
to reformulate the measurement signal in terms of a convolution.

The Wannier states are obtained by representing the position operator $\hat{x}$ 
in the basis of $\hat{h}_0(T_{\rm f})$ and then
diagonalizing it. Afterwards, we set a band limit $\alpha_{\rm max}$.
Modeling the detection efficiencies $\eta_\alpha$ for energetically high lying bands, however, 
is more involved as these depend on both the tunneling and fluorescence imaging time scale.
The tunneling rates grow exponentially with the band index, 
such that they can be divided into tunneling and non-tunneling bands
within the fluorescence imaging time to a good approximation.
For the sake of simplicity, we therefore assume that all (bound) bands lying energetically below $V_{\rm f}$
are detected with unit detection efficiency, meaning \ $\eta_\alpha=1 \ \forall \ \alpha \leq \alpha_{\rm max}$, and 
$\eta_\alpha=0$ for the continuum states,
since atoms in these states are not pinned during the fluorescence imaging.
As a consequence, this model does only give a lower bound 
on the loss in the measurement signal [Eq.\,\eqref{eq:loss}] due to unobserved channels.
We use a lattice containing $L=99$ sites with 33 grid points to resolve each site, 
unless stated otherwise, 
while the potential depth is varied in the range 
$V_{\rm f} \in [50, ... , 300] \ E_{\rm r}$. 

\begin{figure}[h]
	\includegraphics[width=0.5\linewidth]{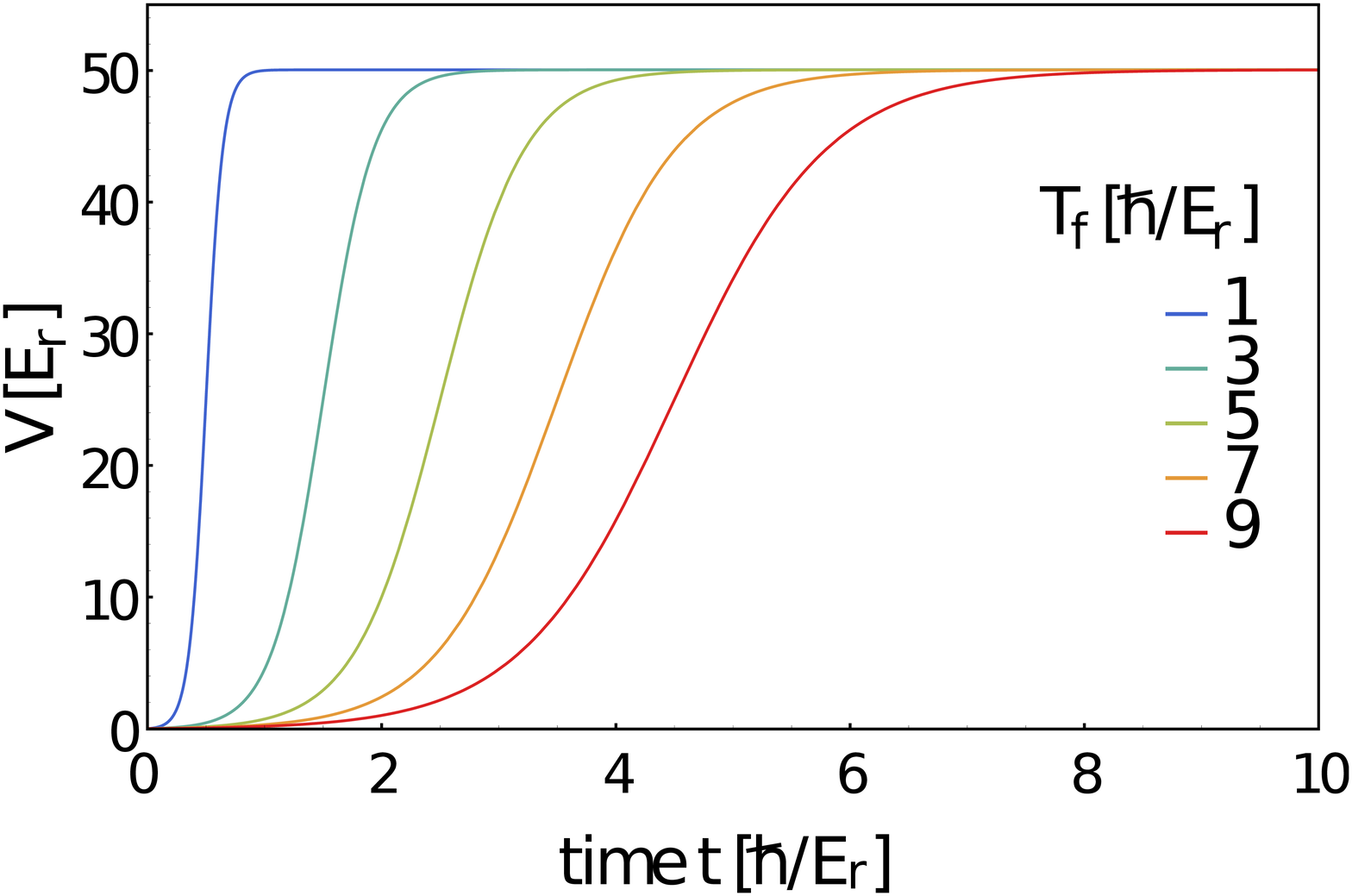}
	\caption{Ramping protocol for $V_{\rm f}=50 E_{\rm r}$, $\eta=10^{-3}$ and varying $T_{\rm f}$.}
	\label{fig:ramping_protocol}
\end{figure}

The back-time propagation of relevant Wannier states is performed
with the Multi-Layer Multi-Configuration Time-Dependent Hartree for bosons (ML-MCTDHB) approach \cite{mctdh1,mctdh2} 
to obtain $|\chi_\alpha\rangle$.
The ramping times cover $T_{\rm f} \in [1, ... , 9] \ \hbar/E_r$ 
and the ramping protocol $V(t)$ is a logistic function of sigmoid form:
\begin{equation}
	V(t) =
	\frac{V_{\rm f}}{1-2\eta} 
	\left(
	\frac{1}
	{1+	\left(   \frac{\eta}{1-\eta}   \right)^{\frac{2 t}{T_{\rm f}}-1}   }
	 - \eta
	 \right),
\end{equation}
with amplitude $V_{\rm max}=V_{\rm f}/(1-2\eta)$, shift $t_0=T_{\rm f}/2$, 
steepness $\frac{1}{\tau}=\frac{2}{T_{\rm f}} 
\ln \left( \frac{1-\eta}{\eta} \right)$ and offset
$V_{\rm off}=V_{\rm max} \eta$. The $\eta$ parameter ensures that $V(T_{\rm f})=V_{\rm f}$ 
does not deviate much from the saturated value $V_{\rm max}$.
With $\eta=10^{-3}$ fixed, $T_{\rm f}$ alone determines the adiabaticity of the ramping protocol
(see Fig.\,\ref{fig:ramping_protocol}).

In the case of an adiabatic preparation of many-body ground states in optical lattices,
such as the bosonic Mott insulator, 
the optimal shape of the ramp function has been extensively discussed~\cite{Gericke2007}.
In contrast, for pinning the distribution on the lattice in quantum gas microscopes, 
simple s-shaped ramps have proven sufficient~\cite{Bakr2010,Sherson2010}. 
We note that in our setting, the dynamics during the ramp will be strongly non-adiabatic in order to avoid a loading of the ground state of the lattice, but freeze the atoms in their original position. 
Therefore, we expect that the precise shape of the ramp should not be important.

\begin{figure}[h]
	\includegraphics[width=0.6\linewidth]{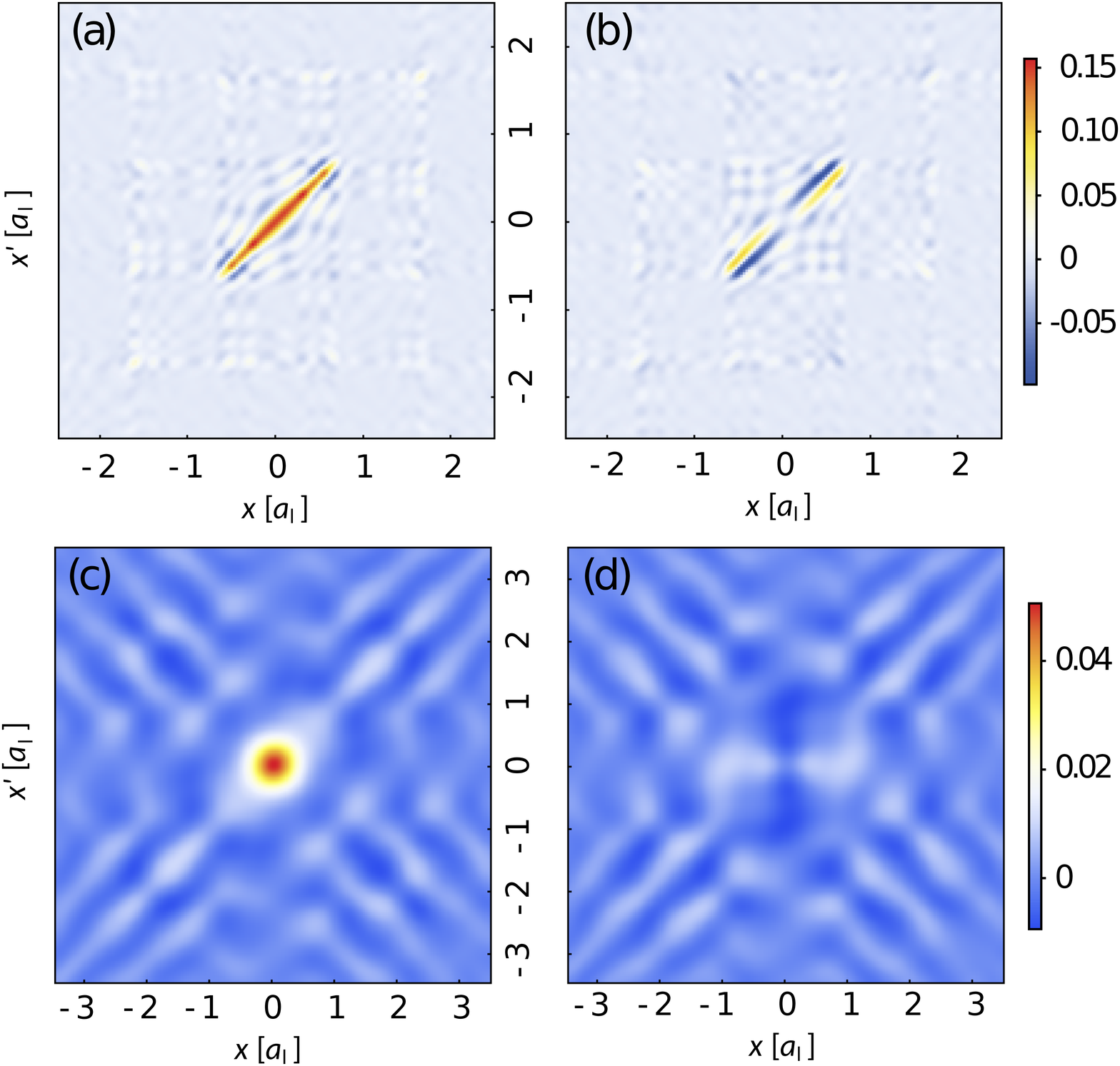}
	\caption{Spatial representation of the qPSF for ramping parameters $V_{\rm f}=200 E_{\rm r}$ and $T_{\rm f}=\hbar/E_{\rm r}$ (a,b) and $V_{\rm f}=50 E_{\rm r}$ and $T_{\rm f}=9 \hbar/E_{\rm r}$ (c,d). The figure shows the real part (a,c) and the imaginary part (b,d).}
	\label{fig:qpsf-comparison}
\end{figure}

We show the real and imaginary part of the spatial representation of the qPSF
$Q(x,x')=\langle x |\hat{Q}|x'\rangle$
for a quick ramp $T_{\rm f}=\hbar/E_{\rm r}$ with a deep lattice $V_{\rm f}=200 E_{\rm r}$ and for a slow ramp $T_{\rm f}=9 \hbar/E_{\rm r}$ with a comparatively shallow lattice $V_{\rm f}=50 E_{\rm r}$ (Fig.\,\ref{fig:qpsf-comparison}). 
In the first case we observe a diagonal pattern in the real part with a
fast decay of the off-diagonal,
while in the second case the real part displays a Gaussian profile with the
imaginary part being
suppressed by an order of magnitude.
Both cases are rather localized around a small region of approximately
$5 a_{\rm l}$.
The diagonal pattern can be induced and enhanced by choosing deeper 
lattices, meaning that higher bands are responsible for this effect, 
although by successively adding bands for the qPSF calculation we found that
approximately only the first half of the bands 
$\alpha_{\rm max}$ is responsible for the pattern formation.
Going to ramp times beyond $T_{\rm f}=9 \hbar/E_{\rm r}$ requires large lattices with more than $L=100$
lattice sites, because the Wannier states, propagated back in time, 
almost reach the boundaries of the grid.

The qPSF has no direct relation to the classical PSF of the imaging system with finite numerical aperture NA, which is used for the fluorescence imaging after the pinning of the atoms. As long as the NA is large enough to allow for a reconstruction of the lattice occupation (typically NA=0.6-0.8), it drops out of the problem. If one repeats the measurement with varying positions of the pinning lattice with respect to the initial system via the displacement by $\varphi$, even the lattice constant $a_{\rm l}$ does not pose a fundamental limit to the resolution. In the numerical examples a sampling with resolution $0.03 a_{\rm l}$ was used and similar relative positioning of $0.1 a_{\rm l}$ between the pinning lattice and further traps were reported experimentally \cite{Weitenberg2011}. The distortion from the dynamics during ramp-up, which is captured by the qPSF and is relevant for structures on the order of $a_{\rm l}$, is therefore the fundamental limitation on the resolution. The deconvolution with the qPSF can then lead to density measurements with a resolution significantly higher than $a_{\rm l}$.

\subsection{Section C: Examples of application}
For the numerical implementation we make use of recoil units $x_{\rm r}=1/k_{\rm l}$, 
$E_{\rm r}=\hbar^2 k_{\rm l}^2 / (2m)$, $T_{\rm r}=\hbar/E_{\rm r}$ 
with the wavenumber $k_{\rm l}=2 \pi / \lambda_{\rm l}$ of the laser beam of wavelength $\lambda_{\rm l}=1064$\,nm 
to create the lattice potential and $m$ being the mass of the trapped particles, here $ ^{87}$Rb. 
The lattice constant is $a_{\rm l}=\lambda_{\rm l}/2$.

\textit{i) Harmonic oscillator (HO) eigenstates}
\begin{equation}
	\psi_n(x)=\frac{1}{\sqrt{n! 2^n}} \frac{1}{\sqrt[4]{\pi}} 
	\sqrt{\frac{1}{a_{\rm ho}}} 
	H_n(\frac{x}{a_{\rm ho}}) 
	exp(-\frac{ x^2 }{2 a_{\rm ho}^2}),
\end{equation}
where $a_{\rm ho}=\sqrt{\frac{\hbar}{m \omega}}$ the harmonic oscillator length, 
$\omega$ the frequency of the trap and $n \in \mathbb{N}_0$ the excitation level. 
To characterize the structure size of the HO modes with respect to the lattice 
we consider the variance of the position operator 
divided by the number of peaks in the density profile
$\sigma_{n}/a_{\rm l}$ with 
$\sigma_{n}=\frac{1}{n+1} (\langle \psi_n |x^2| \psi_n\rangle -\langle \psi_n|x|\psi_n \rangle^2)^{1/2}$.

\textit{ii) Dark soliton}
We prepare a dark soliton within the mean-field approximation 
placed in a reflection-symmetric box with an extension $L_{\rm b}$ 
smaller than that of the pinning lattice $L_{\rm l}=L  a_{\rm l}$. 
We position the soliton in the center of the box and 
ensure that it is sufficiently separated from the walls:
\begin{equation}
	\psi(x)= \left\{
		\begin{array}{ll}
		 - c_1 \tanh(\frac{(L_{\rm b}/2+x)}{\sqrt{2} \xi}) & 
		\text{if} -\frac{L_{\rm b}}{2} < x < -\frac{L_{\rm b}}{2} + 10 \xi \\
		 c_2 \tanh(\frac{x}{\sqrt{2} \xi}) & 
		\text{if} -\frac{L_{\rm b}}{2} + 10 \xi < x < \frac{L_{\rm b}}{2} - 10 \xi\\
		 c_3 \tanh( \frac{(L_{\rm b}/2-x)}{\sqrt{2} \xi}) &
		\text{if} \ \frac{L_{\rm b}}{2} - 10 \xi < x < \frac{L_{\rm b}}{2} \\
		\end{array}
\right.,
\end{equation}
where the prefactors $c_i$ are chosen such as 
to ensure the continuity and the normalization
of the wave function, 
$\xi=1/\sqrt{8 \pi \bar{\rho} a_{sc}}$ is the healing length of the condensate, 
$a_{sc}$ the scattering length and $\bar{\rho}$ the constant background density. 
The structure size is chosen as $\sigma_\xi/a_{\rm l}=2\xi/a_{\rm l}$, 
which is approximately the full-width-at-half-maximum (FWHM) of the soliton profile. 
The soliton example is used for training of the filter only.

\textit{iii) Impurity in a Fermi sea}
We put $N$ spin-polarized fermions in a reflection-symmetric box of length $L_{\rm b}<L_{\rm l}$.
A stationary impurity positioned in the middle of the potential acts
as a repulsive delta-potential of infinite strength, 
inducing a density profile of fermions similar to that
of a soliton, but with an oscillatory background. 
The eigenstates have a defined parity:
\begin{eqnarray}
	\psi_j^{even}(x) &=\sqrt{\frac{2}{L_{\rm b}}}
	\sin{\Bigg(\frac{2 \pi j}{L_{\rm b}} |x|\Bigg)},\\
	\psi_j^{odd}(x) &=\sqrt{\frac{2}{L_{\rm b}}}
	\sin{\Bigg(\frac{2 \pi j}{L_{\rm b}} x\Bigg)}.
\end{eqnarray}

The density operator for an even number of fermions is then given by a mixed state
\begin{equation}
	\hat{\rho}_1=\frac{1}{N} 
	\Bigg(
	\sum_{j=1}^{N/2}  |\psi_j^{even} \rangle \langle \psi_j^{even}| +
	\sum_{j=1}^{N/2}  |\psi_j^{odd} \rangle \langle \psi_j^{odd}|
	\Bigg).
\end{equation}

Here, the structure size is assigned by an average extension of a peak 
in the one-body density $\sigma_N/a_{\rm l}=(L_{\rm b}/N)/a_{\rm l}$.

\textit{iv) Two bosons with infinite repulsion in HO}
The highly correlated problem of two bosons trapped in a harmonic trap and 
interacting with each other via a delta-potential of infinite strength 
can be solved analytically in the relative frame \cite{Busch1998}.
By transforming the solution back into the laboratory frame
and tracing out one of the coordinates 
one obtains the following one-body density matrix:
\begin{equation}
	\rho_1(x,y)= \frac{b^3}{\pi} e^{-0.5 b^2 (x^2+y^2)} \\
	\left\{
		\begin{array}{ll}
		g(x,y) & 
		\text{if} \ x < y \\
		g(y,x) & 
		\text{if} \ y < x \\
		\end{array}
	\right.\textrm{, with}
\end{equation}
\begin{equation}
	g(x,y)=\sqrt{\pi} \Big(xy+\frac{1}{2 b^2} \Big) (\text{erf}(bx)-\text{erf}(by)+1) + \frac{y}{b} e^{-b^2 x^2} - \frac{x}{b} e^{-b^2 y^2}
\end{equation}
and $b=1/a_{\rm ho}$. 
The correlated two-body system requires the full $2D$ convolution
to create the signal, 
which is cumbersome to achieve on a large grid with fine resolution.
So we consider very large trapping frequencies 
and reduce the grid to $L=33$ lattice sites. The structure size is defined similar to
the HO case: $\sigma/a_{\rm l}$ with 
$\sigma^2= \int{\rm d}x \rho(x) x^2 - (\int{\rm d}x \rho(x) x)^2$.

\subsection{Section D: Simulation of the measurement signal}

In the most general formulation the distorted signal $s(z)$ can be obtained
directly via a $2D$ convolution of the one-body density matrix $\rho_1(x,x')$
of the initially prepared system (Section C) with the kernel $Q(x,x')$ (Section B):
\begin{equation}
	s(z)= Tr \{ \hat{T}_z\hat{\pi}\hat{Q}\hat{\pi}\hat{T}_z^{\dagger}\hat{\rho}_1 \}=
	(\rho_1*Q)(z,z) = \int \int {\rm d}x {\rm d}y \ \rho_1(x,y) Q(z-x,z-y).
	\label{eq:signal_operators}
\end{equation}

However, the $(L \cdot 33)\times(L \cdot 33)$ matrices lead 
to approximately $(L \cdot 33)^4$ numerical
operations, 
which renders the direct calculation inefficient for larger grids.
One way to circumvent this issue 
would be to make a smaller support for
the density by confining it more tightly 
and for the filter by defining a cutoff,
when the amplitudes drop below a certain value.
Here, we just verified that a spacing $\Delta x = (1/33) a_{\rm l}$
provides converged signals by doubling the site resolution.

Independently of the above statements we can reduce the numerical effort
to $\propto(L \cdot 33)^2$, namely for weakly correlated systems 
the spectrally-decomposed one-body density operator has
a finite number of natural populations $\lambda_\gamma$ with considerable weight:
\begin{equation}
	\hat{\rho}_1=
	\sum_{\gamma=1}^{\gamma_{\rm max}}\,\lambda_\gamma\,|\phi_\gamma \rangle\!\langle \phi_\gamma|,
\end{equation}
with $|\phi_\gamma \rangle$ natural orbitals. Inserting this relation
and additionally the expansion of the qPSF into Eq.\,\eqref{eq:signal_operators} we obtain the signal as a sum
of 1D convolutions of the natural orbitals $\phi_{\gamma}$ 
with the 'band' filters $\chi^*_\alpha$:
\begin{equation}
	s(z)=
	\sum_{\gamma=1}^{\gamma_{\rm max}}\,
	\sum_{\alpha=1}^{\alpha_{\rm max}}\,\lambda_\gamma \eta_\alpha
	\Big| (\chi^*_\alpha * \phi_{\gamma})(x) \Big|^2.
\end{equation}

There is another important point worth mentioning, namely the padding. 
Since we are working with finite systems, a convolved function spans 
a larger region than the input functions. 
Thus, we need to provide values for chosen densities outside the grid and padding 
with zeroes is the most natural choice for trapped systems, while periodic padding would be suitable for ring geometries. 
Also, we ensure that the distortion of the signal does not reach the boundaries of the grid.

For a given lattice realization, meaning fixed $V_{\rm f}$, $T_{\rm f}$ and phase $\varphi$,
the signal $s(z=i \cdot a_{\rm l} + \varphi)$ with spatial sampling period $a_{\rm l}$ 
sums up to unity only when $\eta_\alpha=1 \ \forall \ \alpha$, because 
$\hat{R}^\dagger_{i;\varphi} \hat{R}_{i;\varphi}$ then forms a positive operator-valued measure.
In our case, neglecting continuum states lying energetically above $V_{\rm f}$ results
in a particle loss $\Omega$. 
In other words, $\Omega=1-\sum_{i=1}^L s(i \cdot a_{\rm l} + \varphi)$ 
is the probability of finding a particle in none of the sites, but in the unobserved channels. 
Averaging over multiple lattice realizations $\varphi \in \{0,...,\pi\}$ 
we can estimate the mean particle loss $\bar{\Omega}$ expected for the given
pre-measurement reduced one-body density $\rho_1$:
\begin{equation}
	\bar{\Omega}(V_{\rm f},T_{\rm f},\rho_1) = 
	1- \frac{1}{\pi}\int_0^\pi{\rm d}\varphi \,\sum_{i=1}^L s({i \cdot a_{\rm l} +\varphi}).
	\label{eq:loss}
\end{equation}

In Table \ref{tab:losses} we show the average loss $\bar{\Omega}$ for densities and ramp-up parameters 
discussed in the main text, which is indeed very small and has a tendency to decrease for
larger structures.

\begin{table}[h]
\begin{center}
  \resizebox{.4\textwidth}{!}{
  \begin{tabular}{  c | l | c | c | c }
    \hline
    \hline
    \multicolumn{1}{c}{} & \multicolumn{1}{c|}{} 
    & \multicolumn{3}{c}{structure size $\sigma/a_l$} \\
    \cline{3-5}
    \multicolumn{1}{c}{} & \multicolumn{1}{c|}{} & $0.5$ & $1.0$ & $2.0$ \\
    \cline{1-5}
    \multicolumn{1}{l|}{} & HO $n=10$ &
    $0.030$     & 
    $0.014$     & 
    $0.013$ \\ 
    \multicolumn{1}{l|}{$\rho_1$} & two-Bosons &
    $0.021$ & 
    $0.015$     & 
    $0.014$      \\
    \multicolumn{1}{l|}{} & Fermi polaron &
    $0.033$     & 
    $0.031$     & 
    $0.017$ \\ 
    \hline
    \hline
  \end{tabular}
  }
\end{center}
\caption{Average particle loss $\bar{\Omega}(V_{\rm f},T_{\rm f},\rho_1)$ for the densities, discussed in the main text, and different structure sizes $\sigma$ (see Section C) relative to the lattice spacing. The ramp-up parameters are $V_f=200 E_r$ and $T_f=\hbar/E_r$.}
\label{tab:losses}
\end{table}

\subsection{Section E: Multi-frame filter}

Our starting point is Eq.\,(6) of the main text. 
First, we generate a batch of signals $s^{(k)}(z)$ with $k=1,...,n_t$ by the full $2D$ convolution
of chosen one-body density matrices $\rho_{1}^{(k)}(z,z')$ with the qPSF $Q(z,z')$.
Then we estimate each signal $s^{(k)}(z)$ as a 1D convolution
of the corresponding densities $\rho^{(k)}(z)=\rho_{1}^{(k)}(z,z)$ with the same 
density-independent filter $q(z)$: 
\begin{eqnarray}
	s^{(k)}(z)
	= (\rho_{1}^{(k)} * Q )(z,z) 
	\approx (\rho^{(k)} * q )(z).
\end{eqnarray}

Further, we assume $q$ to be space invariant, but otherwise
no priors  will be imposed, because it has no physical interpretation 
and is rather a mathematical tool.
Thus, while the densities of the training set vary from signal to signal, the same
filter $q$ is common to all signals. Each of them provides additional
information on $q$, thereby restricting the space of possible solutions.

As to the choice of training samples 
we create a random selection of five soliton samples with different 
extension $\xi \in [10,...,40] \times 10^{-8}$\,m
in a box $L_{\rm b} \in [3/4,...,10/12]L_{\rm l}$
as well as a random selection of five HO samples 
with trapping frequency $\omega \in [200,...,800] \times 2\pi$\,Hz
and excitation level $n \in \{1,...,10\}$.
We explicitly don't include correlated examples from Section C
to later test the filter $q$ on unknown signals. 

In the next step we define a total loss function $\mathcal{L}$, which
describes a deviation between the true signals $s^{(k)}$ and
their approximations $\rho^{(k)} * q$, a least squares problem:
\begin{equation}
	\underset{q}{\min} \ \mathcal{L} (q)
	=\underset{q}{\min} \ 
	\frac{1}{n_t} \sum_{k=1}^{n_t} \int{\rm d}z|s^{(k)}(z) - (\rho^{(k)} * q)(z)|^2 
	\approx\underset{\mathbf{q}}{\min} \ 
	\frac{1}{n_t} \sum_{k=1}^{n_t} |\mathbf{s}^{(k)} - \mathbf{A}^{(k)} \mathbf{q}|^2,
\end{equation}
where in the last step we switch to a numerical grid with $\mathbf{s}^{(k)}$ and $\mathbf{q}$ being
$(L\cdot 33)$-dimensional vectors and
$\mathbf{A}^{(k)}$ denoting a $(L\cdot 33 \times L\cdot 33)$ Toeplitz matrix, which
represents a 1d discrete convolution with zero padding and limited support.

To find the filter $q$, that is more likely to have created the
observed distortions in the signals,
we perform the gradient descent algorithm in batch mode,
meaning that we take into account all the frames simultaneously.
We find that a small amount of samples is sufficient
to obtain a well-performing filter. Thus, we do not need to resort
to more memory-efficient optimization algorithms such as stochastic or mini-batch gradient descent.

In each iteration step $m \in \mathbb{N}_0$ the filter is updated such that we follow a path towards the minimum of $\mathcal{L}$
by taking a direction of negative gradient $\nabla \mathcal{L}$. As initial guess we take the diagonal of the qPSF 
$\mathbf{q}_0=Q(x,x)$ and then iterate
\begin{equation}
	\mathbf{q}_{m+1}=\mathbf{q}_m - \beta \nabla \mathcal{L}(\mathbf{q}_m),
\end{equation}
where the gradient of the loss function reads:
\begin{equation}
	 \nabla \mathcal{L}(\mathbf{q}_m)
	 = \frac{1}{n_t} \sum_{k=1}^{n_t} 2 \mathbf{A}^{(k)^T}(\mathbf{A}^{(k)} \mathbf{q}_m - \mathbf{s}^{(k)}).
\end{equation}

The step size or learning rate $\beta$ can be optimally 
calculated (accurate line search) for each iteration step as
\begin{align}
	\beta &=\underset{\beta}{\text{argmin}} \ \mathcal{L} (\mathbf{q}_m - \beta \nabla L(\mathbf{q}_m))
	\nonumber \\
	&=\frac{1}{2}
	\frac{\sum_k (\mathbf{q}_m^T \mathbf{A}^{(k)^T}-\mathbf{s}^{(k)^T}) \mathbf{A}^{(k)} \mathbf{A}^{(k)^T} 
	(\mathbf{A}^{(k)} \mathbf{q}_m-\mathbf{s}^{(k)})}
	{\sum_k (\mathbf{q}_m^T \mathbf{A}^{(k)^T}-\mathbf{s}^{(k)^T}) 
	\mathbf{A}^{(k)} \mathbf{A}^{(k)^T} \mathbf{A}^{(k)} \mathbf{A}^{(k)^T} (\mathbf{A}^{(k)} \mathbf{q}_m-\mathbf{s}^{(k)})}.
\end{align}
Finally, we iterate until the relative change in the total loss function reaches some threshold.

\subsection{Section F: Deconvolution}
During the image acquisition by microscopes in molecular biology \cite{Wallace2001} 
or telescopes in astronomy \cite{Schulz1993}
multiple degradation sources can distort the true form of the object:
noise, scatter, glare and blur.
The blur, caused by the passage of light through
the  imaging system, leads to a non-random light redistribution and
poses a fundamental limitation to the imaging system.
The recorded image is usually modeled as
a convolution of the object with a filter, also known as point spread function (PSF).
There exists a variety of methods to reverse this process and retain
the original object, called deblurring or deconvolution algorithms \cite{Hansen2006}.
They can be classified as inverse (Wiener-Deconvolution \cite{wiener1949}) or iterative 
(Van-Cittert \cite{van_cittert}, Lucy-Richardson \cite{Lucy1974,Richardson1972}, Steepest Descent \cite{steepest_descent}); 
with prior knowledge of the filter (non-blind deconvolution) 
or completely unknown (blind deconvolution \cite{Ayers1988}); 
imposing priors such as non-negativity and smoothness or without them;
modeling potential noise sources or neglecting them;
using a single frame or a batch of sampled frames (multi-frame deconvolution \cite{Sheppard1998}).

For our case we require a package, which implements an iterative algorithm,
as they are more stable and provide a better restoration of
degraded resolution, although at the cost of longer computation times.
Since we obtained the filter in Section E it should be non-blind.
The density we are trying to reconstruct is positive, so 
a non-negativity constraint is a must, but otherwise
no pre-filtering is necessary.
We also neglect all sources of noise and 
the measurement signal is considered as a single frame.

\textsc{wolfram mathematica 10.4} provides two suitable algorithms for this purpose:
Lucy-Richardson and Steepest Descent. The output is a positive function.
We iterate until converged, disable the pre-filtering and don't include any noise.
Steepest Descent proves to be more reliable than Lucy-Richardson and
the reconstruction is usually of a better quality.

\end{document}